\documentclass{elsart5p}
\usepackage{amssymb, url}
\usepackage{graphicx}
 
\begin{document}

\begin{frontmatter}
\title{Neutron Scattering Facility for Characterization of CRESST and EURECA Detectors at mK Temperatures}

\author[tu]{J.-C.~Lanfranchi}, \author[tu]{C.~Ciemniak}, \author[tu]{C.~Coppi}, \author[tu]{F.~von~Feilitzsch},  \author[tu]{A.~G\"utlein}, \author[tu]{H.~Hagn}, \author[tu]{C.~Isaila},  \author[tb]{J.~Jochum}, \author[tb]{M.~Kimmerle},  \author[tu]{S.~Pfister}, \author[tu]{W.~Potzel}, \author[qu]{W.~Rau}, \author[tu]{S.~Roth}, \author[tb]{K.~Rottler}, \author[tb]{C.~Sailer},  \author[tb]{S.~Scholl},  \author[tb]{I.~Usherov} and \author[tu,ww]{W.~Westphal}

\address[tu]{Physik-Department E15, Technische Universit\"at M\"unchen, D-85747 Garching, Germany}
\address[tb]{Eberhard-Karls-Universit\"at T\"ubingen, D-72076 T\"ubingen, Germany}
\address[qu]{Department of Physics, Queen's University, Ontario K7L 3N6, Canada}
\address[ww]{Deceased}

\begin{abstract}
CRESST (Cryogenic Rare Event Search with Superconducting Thermometers) is an experiment located at the Gran Sasso underground laboratory and aimed at the direct detection of dark matter in the form of WIMPs. The setup has just completed a one year commissioning run in 2007 and is presently starting a physics run with an increased target mass.  Scintillating $\mathrm{CaWO_4}$ single crystals, operated at temperatures of a few millikelvin, are used as target to detect the tiny nuclear recoil induced by a WIMP. 
The powerful background identification and rejection of $\alpha$, e$^{-}$ and $\gamma$ events is realized via the simultaneous measurement of a phonon and a scintillation signal generated in the $\mathrm{CaWO_4}$ crystal. However, neutrons could still be misidentified as a WIMP signature. Therefore, a detailed understanding of the individual recoil behaviour in terms of phonon generation and scintillation light emission due to scattering on Ca, O or W nuclei, respectively, is mandatory. The only setup which allows to perform such measurements at the operating temperature of the CRESST detectors has been installed at the Maier-Leibnitz-Accelerator Laboratory in Garching and is presently being commissioned. The design of this neutron scattering facility is such that it can also be used for other target materials, e.g. $\mathrm{ZnWO_4}$, $\mathrm{PbWO_4}$ and others as foreseen in the framework of the future multitarget tonne-scale experiment EURECA (European Underground Rare Event Calorimeter Array).
\end{abstract}

\end{frontmatter}

\section{Introduction}
CRESST (Cryogenic Rare Event Search with Superconducting Thermometers) is located at a depth of 3600~m.w.e. (meter water equivalent) in Hall A of the Laboratori Nazionali del Gran Sasso (L.N.G.S.), Italy.
The experiment is aimed at the direct detection of WIMPs with cryogenic 
detectors consisting of 300~g $\mathrm{CaWO_4}$ crystals equipped with tungsten transition edge sensors (TESs). These target crystals are additionally monitored by cryogenic light detectors to measure the amount of light emitted per particle interaction. The simultaneous measurement of the heat and the scintillation signal of the $\mathrm{CaWO_4}$ crystal allows a highly efficient background rejection \cite{meunier99, angloher2005}.\\
After a major upgrade phase (CRESST-I to CRESST-II), the cryostat can now accommodate 10~kg of total target mass in the detector carousel. CRESST-II was also upgraded concerning the shielding which now comprises an additional 45~cm polyethylene shield against neutrons and an active muon veto.
With this current setup, CRESST should be able to reach a sensitivity for the spin-independent WIMP-nucleon cross section in the range of 10$^{-8}$~pb.

$\mathrm{CaWO_4}$ contains a heavy nucleus (tungsten), which makes it a good target 
for spin-independent coherent WIMP interactions \cite{angloher2005}. 
The crystals used in the experiment are cylindrical $\mathrm{CaWO_4}$ single crystals of 40~mm diameter and 40~mm height with a mass of  $\sim$300~g equipped with tungsten transition edge sensors (TESs) directly evaporated onto them. 
With the TES the phonon signal induced by an event is detected, delivering the 
primary information about the deposited energy. This part of the detector module is referred to as the Òphonon detectorÓ. 

For the detection of the scintillation light, cryogenic light detectors also equipped with tungsten TESs are used. 
The first generation of light detectors for CRESST consisted of 30x30x0.4~mm$^{3}$ silicon wafers to absorb the scintillation light emitted by the $\mathrm{CaWO_4}$ crystal ($\sim$420~nm).  The second generation of light detectors is made of sapphire discs of 40~mm diameter coated with silicon on one side (ÒSOSÓ: silicon on sapphire). The absorber crystals were changed from silicon to SOS in order to benefit from the excellent phonon transportation properties in sapphire.

\begin{figure}\centering
\includegraphics[width=.5\textwidth]{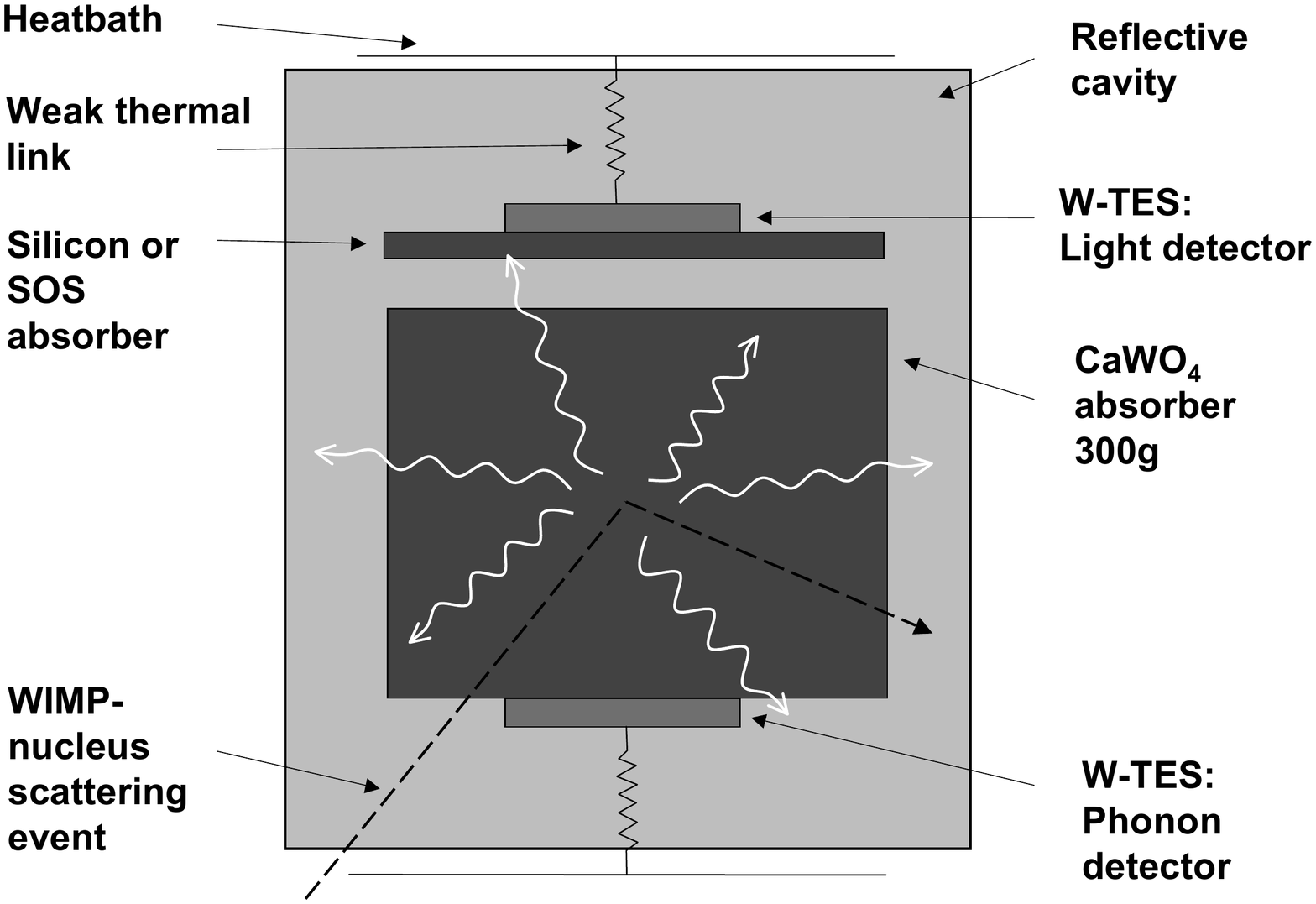}
\caption{Schematic setup of a CRESST detector module. The module consists of two independent detectors: one phonon channel and one light channel. Both detectors are located in a reflective housing.}\label{fig: detector module}
\end{figure}

Phonon and light detectors are mounted together in a copper housing. The inner surfaces are covered with a highly reflective scintillating polymeric foil for efficient light collection. Fig.~\ref{fig: detector module} depicts a schematic drawing of one such detector module. 

The light output of an event is characterized by its Òlight yieldÓ Y defined as 
Y = Signal Amplitude in Light Detector/Signal Amplitude in Phonon Detector. 
Y is usually normalized to 1 for electron recoil events. For other event classes, 
like alphas or nuclear recoils from neutron or WIMP interactions, the light 
yield is reduced. The reduction of the light yield is expressed by the quenching factor Q = 1/ Y .

Most neutron recoil events in the relevant energy window between 10 keV and 40 keV 
are due to scattering off oxygen. On the other hand, WIMP events 
in this energy window are expected to be mainly scatterings off tungsten 
nuclei in the $\mathrm{CaWO_4}$ crystal. Thus, the identification of the 
recoiling nucleus is an extremely helpful tool to discriminate the neutron background from 
a possible WIMP signal. This can be achieved by measuring the quenching 
of the scintillation light with respect to the recoiling nucleus. There are several 
approaches to measure the quenching factors for the elements in $\mathrm{CaWO_4}$ \cite{ninkovic2006, jagemann2006, coppi2006, bavykina}. In this article a neutron scattering facility, set up at the Maier-Leibnitz-Accelerator-Laboratory (MLL), using a pulsed monochromatic neutron beam of 11 MeV to measure quenching factors at mK temperatures is described.\\

The facility described in this article not only is of relevance for CRESST but also for the planned EURECA (European Underground Rare Event Calorimeter Array) experiment \cite{kraus07}. EURECA will be the future European dark matter experiment using low-temperature detectors. It will combine the efforts of the present CRESST, EDELWEISS (Experience pour DEtecter Les WIMPs en Site Souterrain) \cite{edelweiss05} and ROSEBUD (Rare Objects SEarch with Bolometers UndergrounD) \cite{rosebud05} experiments in a new experimental hall in the Laboratoire Souterrain de Modane (LSM). Once fully set up, EURECA will exhibit a total mass on the tonne scale using multiple target materials.

\section{Experimental Setup}\label{setup}

\begin{figure}\centering
\includegraphics[width=.5\textwidth]{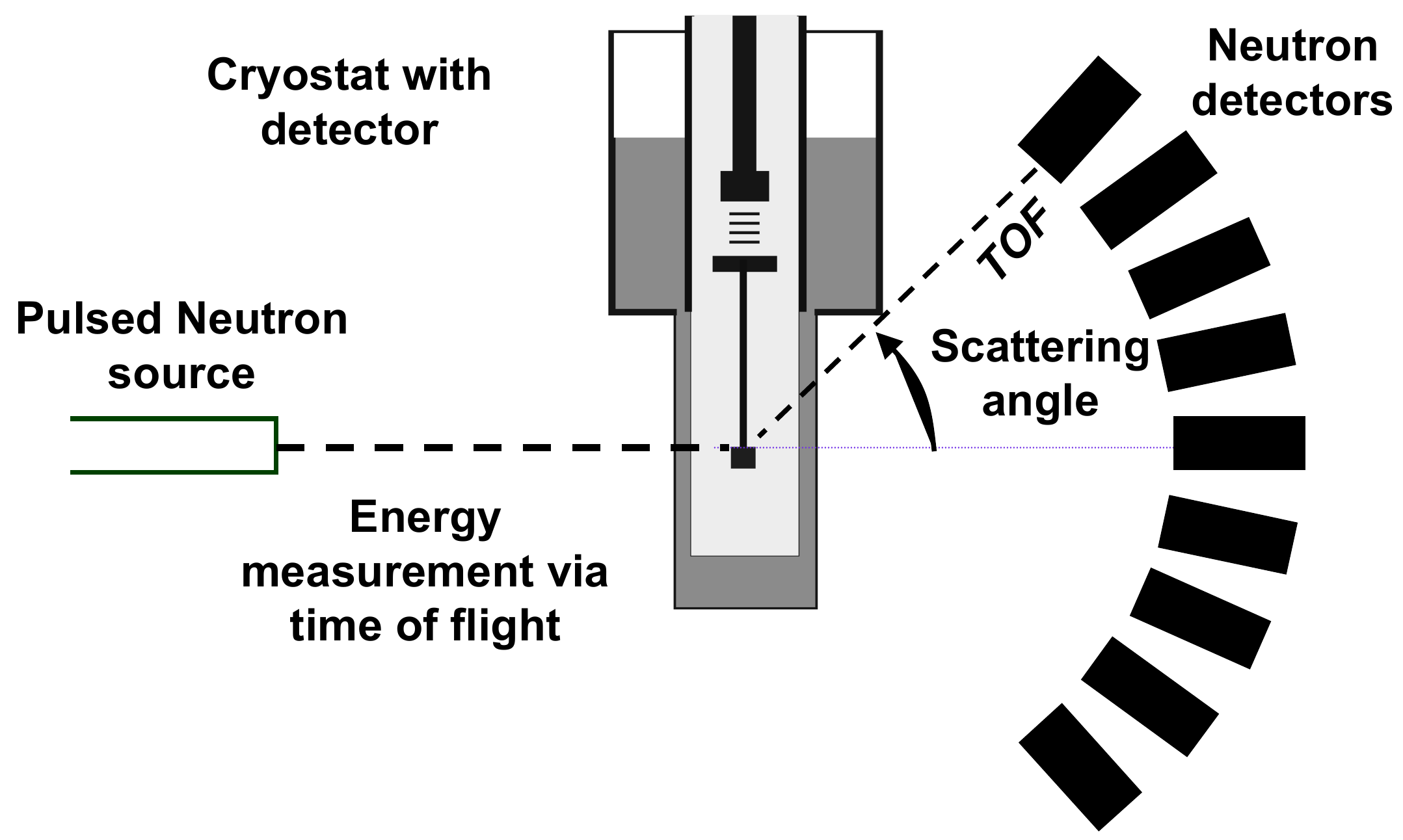}
\caption{Schematic of the neutron scattering facility setup. It consists of a pulsed neutron source, a central detector consisting of a $\mathrm{CaWO_4}$ cryogenic detector placed inside a cryostat and an array of $\sim$40 neutron detectors.
The dewar of the cryostat is designed such that a minimum amount of liquid helium is present in the near surroundings of the $\mathrm{CaWO_4}$ crystal to avoid additional scattering of neutrons on the helium.}\label{fig: nsfsetup}
\end{figure}

For a good determination of the quenching factors in a neutron scattering 
experiment it is desirable that the signals for the three different elements
 form distinct populations in the data. This can be achieved by performing 
the scattering with monoenergetic neutrons and selecting a single scattering 
angle under which a time of flight (TOF) measurement is carried out. In this way, the kinematics of the scattering reaction is fixed and the nuclei can be discriminated via the recoil energy. 
The setup of the neutron scattering facility is depicted in Fig.~\ref{fig: nsfsetup}. A similar type of experiment \cite{sicane} was also performed by the EDELWEISS collaboration \cite{edelweiss05}.

The neutrons are produced via the reaction $^{1}$H($^{11}$B, n)$^{11}$C. 
Boron ions with a kinetic energy of 60 MeV from the accelerator are directed 
onto a hydrogen gas target, producing neutrons with an energy of 11 MeV \cite{jagemann2006}. The pulsing of the boron beam provides the reference time for the time-of-flight measurement of the neutrons. 
The neutron scattering facility consists of a central detector unit containing the scintillator crystal to be investigated and a set of $\sim$40 neutron detectors placed at a fixed angle with respect to the beam and the central 
detector, thus defining the scattering geometry. The neutron detectors consist of chambers with NE213 liquid scintillator read out by PMTs. NE213 allows a distinction between neutron and gamma events via pulse shape discrimination.

\subsection{Cryostat}

The central detector is cooled to its operational temperature in a $^{3}$He/$^{4}$He dilution refrigerator. A KELVINOX400 with a cooling power of 400~$\mu$W at 100~mK and a base temperature of $\sim$10~mK was installed in Hall 2 of the MLL. The cryostat uses a special dewar that is designed such that only a minimum amount of liquid $^{4}$He is present around the detector (see Fig.~\ref{fig: nsfsetup}) in order to suppress parasitic neutron scattering on the helium. The cryostat has been equipped with 2 SQUID readout channels for the central cryodetector. The cryostat can be cooled from room temperature to $\sim$10~mK in 
$\sim$1 day and exhibits an extremely stable base temperature over time. Special care was given to decouple the cryostat as well as possible from the mechanical surroundings in order to avoid microphonics.

\begin{figure}\centering
\includegraphics[width=.5\textwidth]{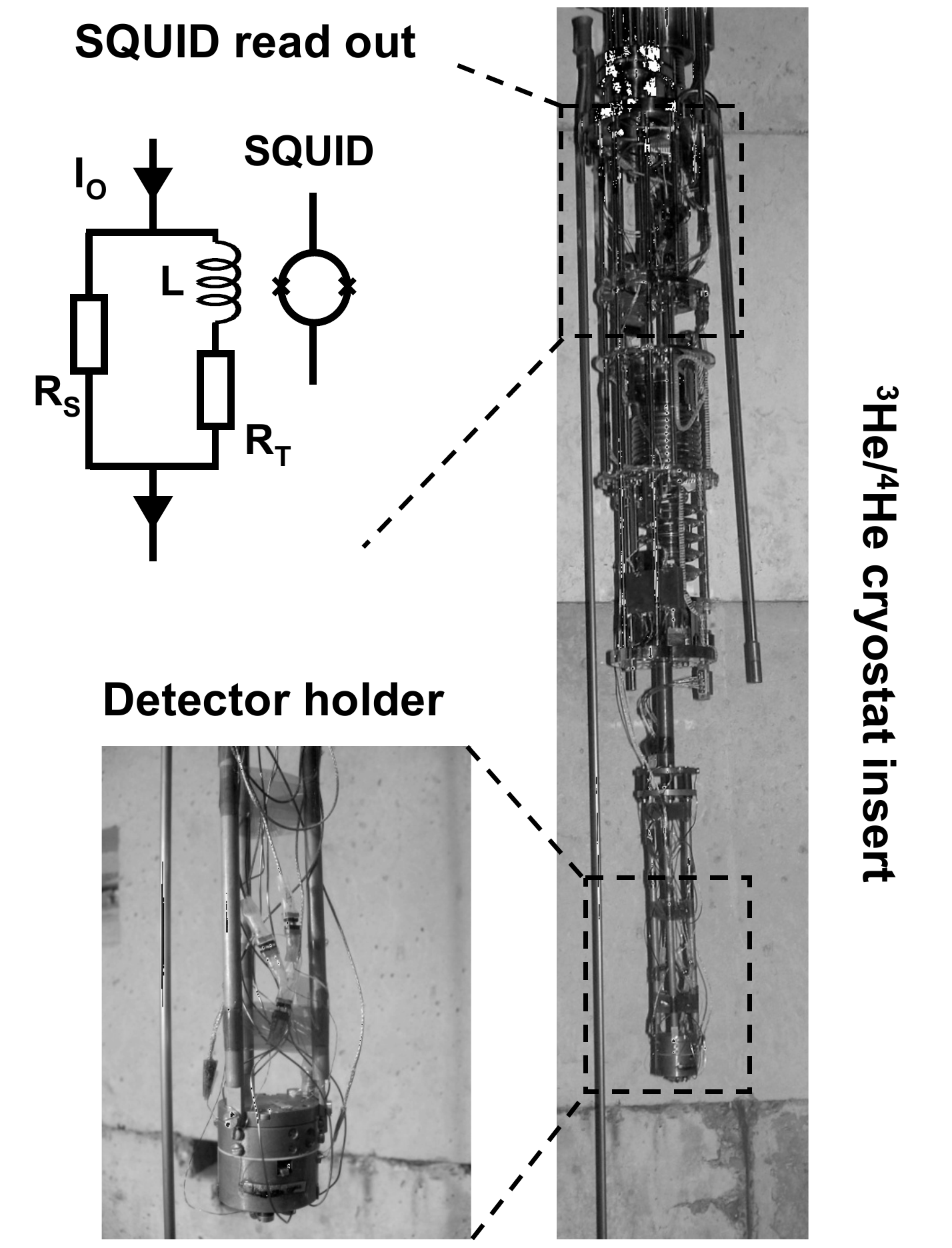}
\caption{Pictures of the inside of the neutron scattering facility's cryostat of the KELVINOX400 (OXFORD Instruments) type. The principle of the SQUID readout system is also indicated, where R$_T$ is the TES's resistance and R$_S$ a reference (shunt) resistance; I$_0$ is the constant bias current.}\label{fig: nsfcryostatinside}
\end{figure}

Fig.~\ref{fig: nsfcryostatinside} shows pictures of the insert of the cryostat and the mounted cryodetector prior to the first beamtime as well as the principle of the SQUID readout design.

\subsection{Dedicated Cryodetector}

A crucial issue in setting up the neutron scattering facility was the development of a dedicated low-temperature detector able to cope with the high count rates. 
 The requirements regarding the detector design for the neutron calibration measurements are different from those of the dark matter experiment.\\ 
Two major aspects had to be considered for the neutron scattering facility detector: 
In the scattering experiment the kinematics are used to identify the recoiling nucleus. Thus, it is important that only a single scattering occurs in the crystal. Therefore, the size of the crystal 
has to be chosen such that one gets as few multiple scattering events as possible, while still having a reasonable event yield for a given neutron flux.
 
 Since the whole calibration experiment is performed above ground with little 
shielding against natural and cosmic radiation and additional events are induced by the calibration sources (gamma, neutrons) high count rates have to be dealt with.
 This count rate is in the order of a few tens per seconds compared to less than one per second in the dark matter experiment. Therefore the detectors have to be designed such that they can handle these 
increased count rates. For details see \cite{phdwolfgangwestphal}. 

\begin{figure}\centering
\includegraphics[width=.5\textwidth]{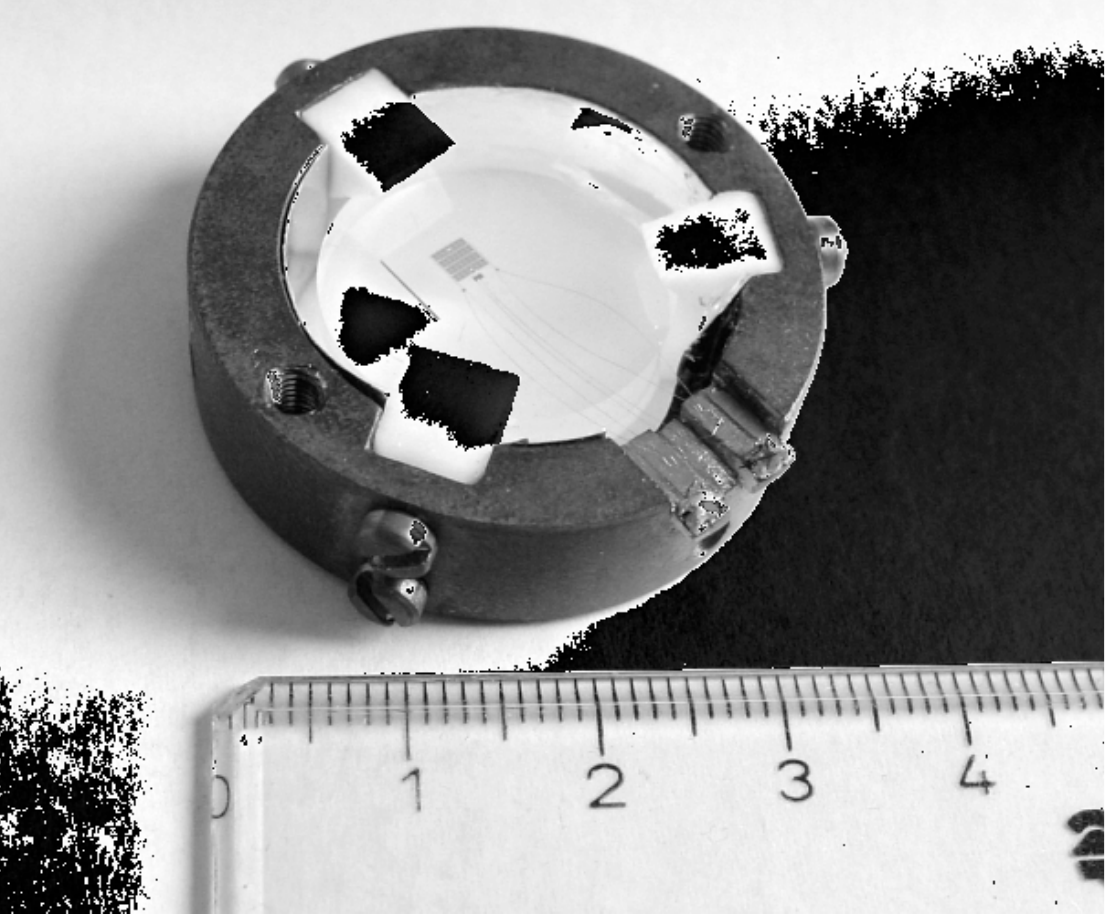}
\caption{Picture of the dedicated phonon detector for high count rates. The 20 mm (diameter) x 5 mm $\mathrm{CaWO_4}$ crystal is placed inside its copper holder. The TES equipped with aluminum phonon collectors can be seen in the middle. }\label{fig: nsfphonondetector}
\end{figure}

The detector which was developed, taking the above mentioned facts into account, consists of  a $\mathrm{CaWO_4}$ cylinder with 20 mm diameter and 5 mm height and is depicted in 
Fig.~\ref{fig: nsfphonondetector}. 

\begin{figure}\centering
\includegraphics[width=.5\textwidth]{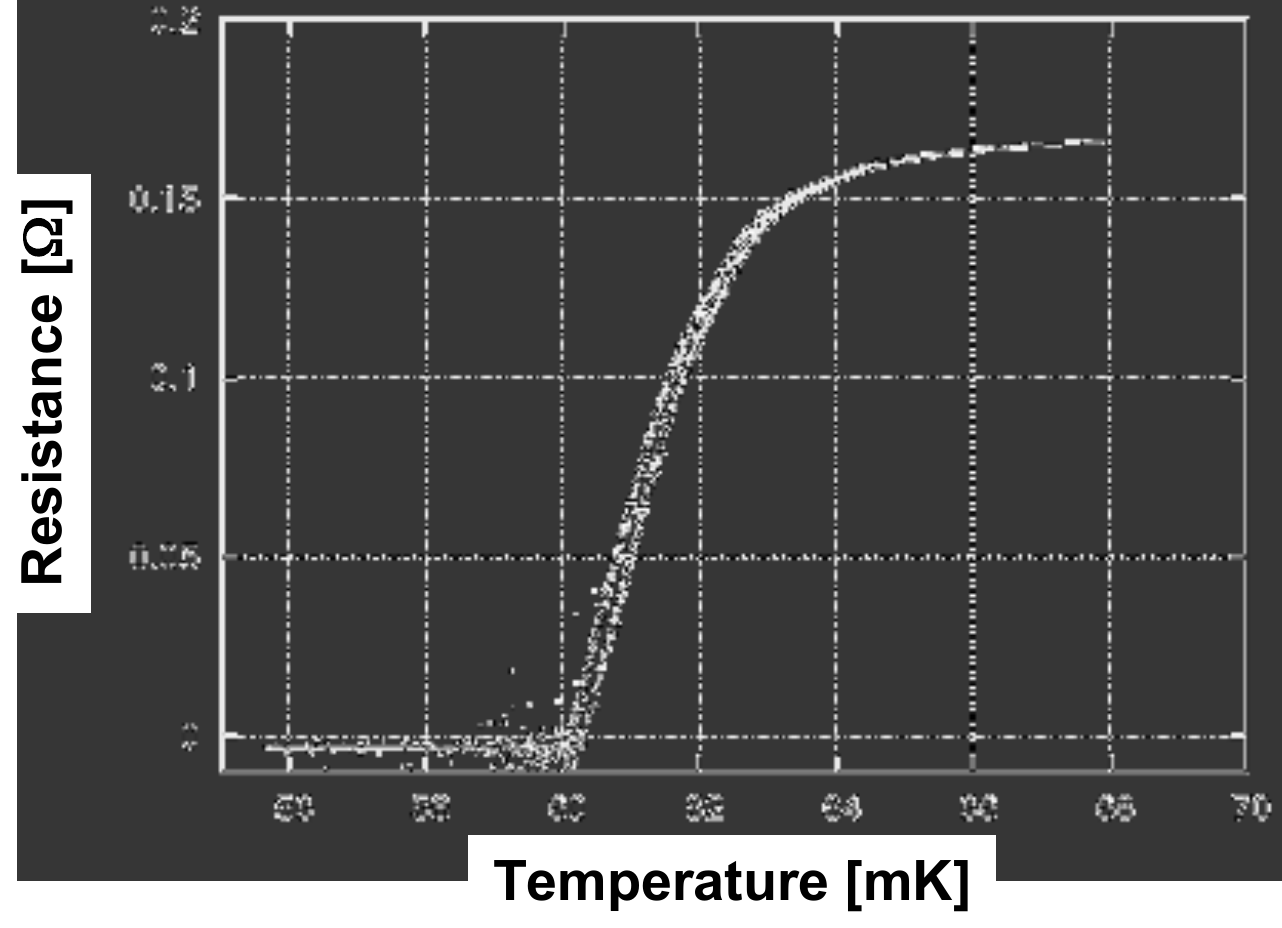}
\caption{Transition curve from the normal to the superconducting state of the phonon detector Cw520. The TES exhibits a highly linear transition between $\sim$60 and 61~mK and a total transition width of only 3 mK. The doubling of the curve (hysteresis) originates from sweeping back and forth through the region of transition.}\label{fig: nsfdetectortransition}
\end{figure}

Also the choice of the TES design necessitated to take several aspects into account. 
First of all, the detector has to be fast enough for the expected count rates in the experiment. This can be achieved by collecting the non-thermal phonons generated in the crystal before they thermalize by introducing additional superconducting phonon collectors \cite{qetforcdmszipdetectors, qetsensor}.

For small signals the measured amplitude at the SQUID output is proportional to the temperature change in the TES. Thus the amplitude for a 
given energy absorbed in the TES scales with the reciprocal heat capacity 
C$_{e}^{-1}$ of the TES.
While a larger sensor area increases the amount of energy collected in 
the TES, the signal amplitude per energy diminishes.  
Furthermore, for the quenching factor measurements a high light yield is crucial. 
It is thus desirable to keep the sensor as small as possible to reduce the risk 
of too large a light loss.

These considerations led to the decision to use a relatively small TES
structure with additional aluminum phonon collectors.

The sensor structure consists of a parallel connection of 28 individual Ir/Au TESs, each 
connected to 10 aluminum phonon collector fins. Each fin has a size of 50x250 $\mu$m$^{2}$,
so that the total phonon collection area is 3.5 mm$^{2}$ . This structure provides a good compromise between the fast phonon collection of a large sensor and a good threshold and small light loss of a small sensor. Fig.~\ref{fig: nsfphonondetector} shows a photograph of the completed detector (Cw520) mounted in its holder. 
Fig.~\ref{fig: nsfdetectortransition} shows the normal-to-superconducting transition curve of the 
sensor. The transition temperature is at about 60 mK. Additionally, the small transition width 
of less than 3 mK and the high linearity of the lower part of the transition 
are ideal prerequisites for a sensitive detector. 

During test measurements the detector showed good performance at low energies with a threshold of 250 eV and a resolution of 340 eV at 5.9 keV. A dynamic range of at least 1.4 MeV was reached in these measurements \cite{phdwolfgangwestphal}.

\section{Preliminary Results}\label{preliminaryresults}

A first set of measurements was performed to commission the neutron scattering facility. For this preliminary phase, it was decided to perform a simplified kind of measurement, by irradiating the cryodetector with a monoenergetic neutron beam from the accelerator, without performing a time-of-flight 
measurement.
The nuclear recoil spectrum for each nucleus will then be continuous, from zero 
up to a maximum energy, which depends on the energy of the neutrons and on the mass of the nucleus. 

\begin{figure}\centering
\includegraphics[width=.5\textwidth]{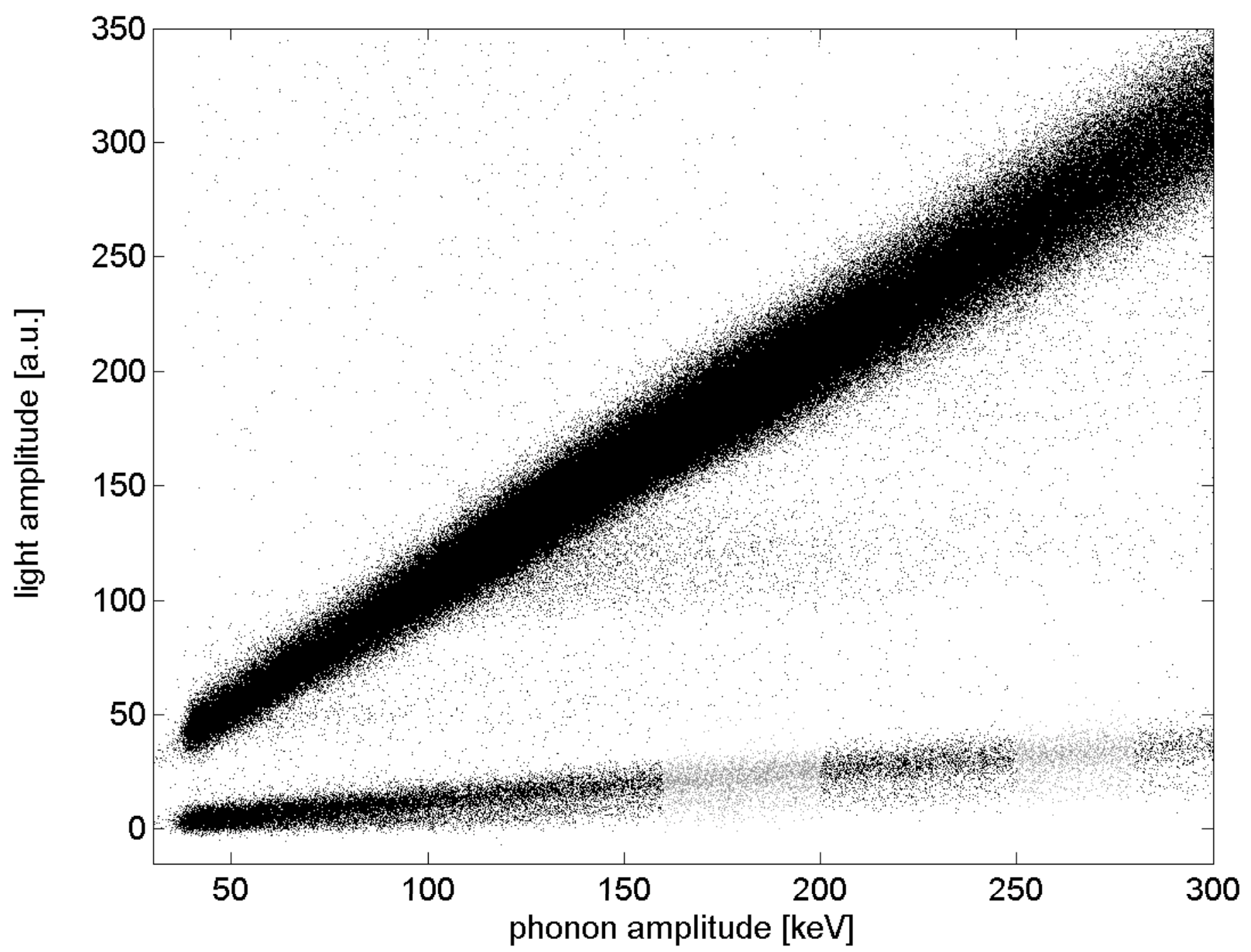}
\caption{Scatter-plot of pulse height in the light detector (in keV electron equivalent) versus pulse height in the phonon detector. The upper band corresponds to electron recoils, the lower band to nuclear recoils. In between the two bands mixed neutron and gamma events are observed originating from inelastic neutron scatterings on tungsten nuclei. Both bands do not join at the low energy part of the spectrum since the trigger threshold was too high during this measurement. The two regions marked in grey from 160 to 200~keV and from 250 to 280~keV in the nuclear recoil band were used to produce the histograms depicted in Fig. \ref{3gauss} and Fig. \ref{2gauss}.}\label{fig: nsf1stbeamtimescatterplot}
\end{figure}

The amplitude of the phonon signal is close to the total energy of the interaction since
more than 95$\%$ \cite{phdwolfgangwestphal} of the energy of an interaction goes into the production of phonons. 
On the other hand, the light output depends on the kind of interacting particles, e.g. electron recoils produce significantly more light than nuclear recoils. Thus, different kinds of events will have different
positions in the light-phonon scatter-plot, forming several bands. In Fig.~\ref{fig: nsf1stbeamtimescatterplot} one can clearly distinguish the electron recoil band (higher light yield) and the nuclear recoil band. 

Moreover, there are three different kinds of nuclei in the crystal, each with a different light yield, so that the nuclear recoil band is actually composed of three bands: the oxygen, the calcium and the tungsten band.

Each class of nuclear recoil events (oxygen, calcium and tungsten) exhibits a Gaussian distribution. Since the difference in the light yield is relatively small, the Gaussians overlap. However, it is possible to fit them and get a value for the yield (Y) and for the quenching factor (QF = 1/Y) of the individual nuclei.

\begin{center}
\begin{figure}
  \includegraphics[width=8.5cm]{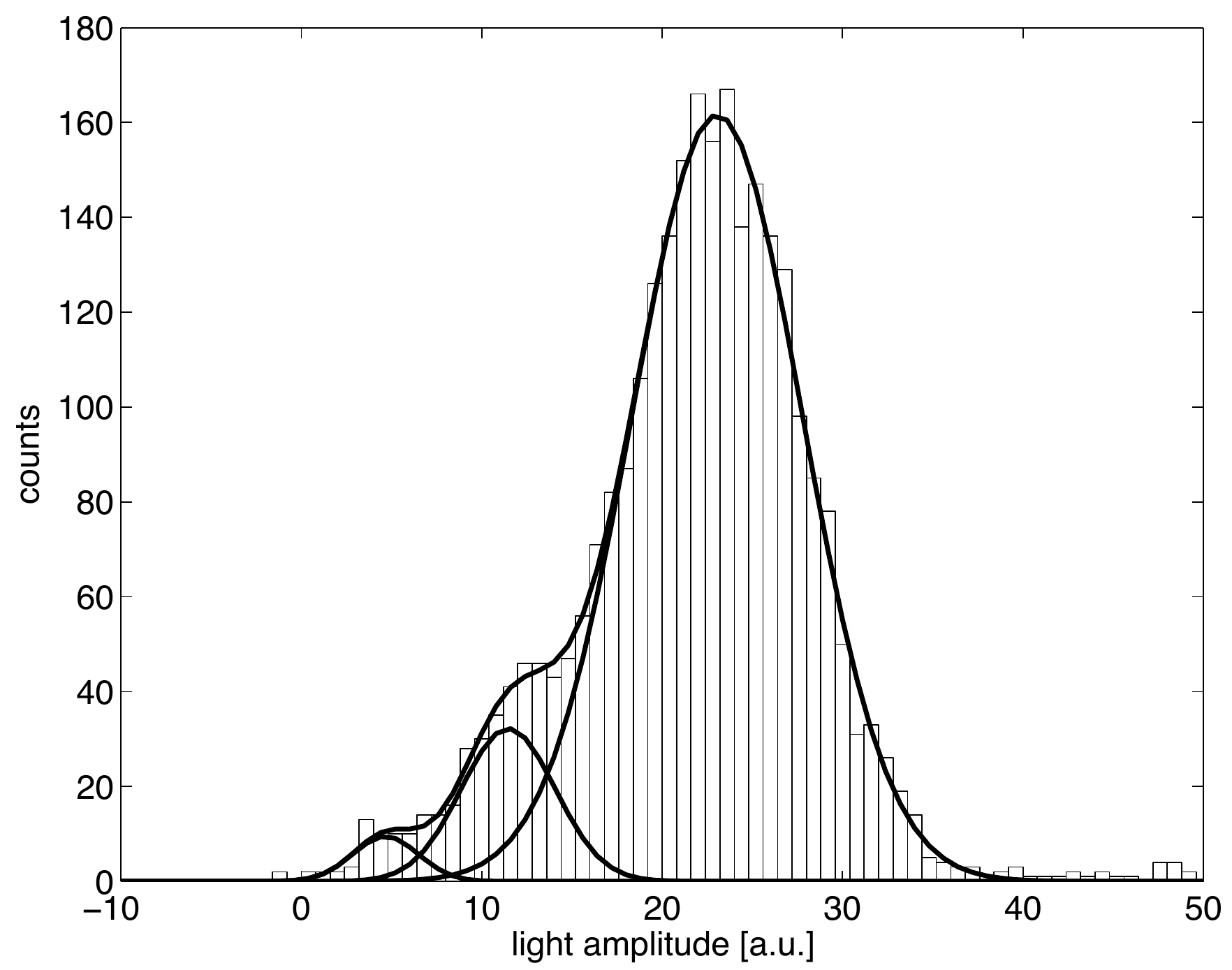} 
  \caption{Histogram of the light amplitude for a section of the nuclear recoil band
(160-200~keV) where contributions from O, Ca and W are expected. A three-Gaussian
fit can be
performed, in order to obtain values for the quenching factors of the three nuclei.}
  \label{3gauss}
\end{figure}
\end{center}

This method can be applied in different energy ranges where the contribution from tungsten recoils is
more or less significant.
In Fig.~\ref{3gauss} the histogram of the light amplitude of nuclear recoil events in the energy interval between 160-200 keV is shown. 

The events depicted in the histogram of Fig.~\ref{3gauss} reflect those marked in grey (160-200~keV) in Fig.~\ref{fig: nsf1stbeamtimescatterplot}. In this case a fit had to be performed with the sum of three Gaussians, one of which (the tungsten one) has a very small amplitude, which makes the fit difficult. The error of the position of the smallest (tungsten) Gaussian is therefore rather big. In Table~\ref{tab_QF1} the results for the QFs (160-200~keV) are summarized.

\begin{center}
\begin{figure}
  \includegraphics[width=8.5cm]{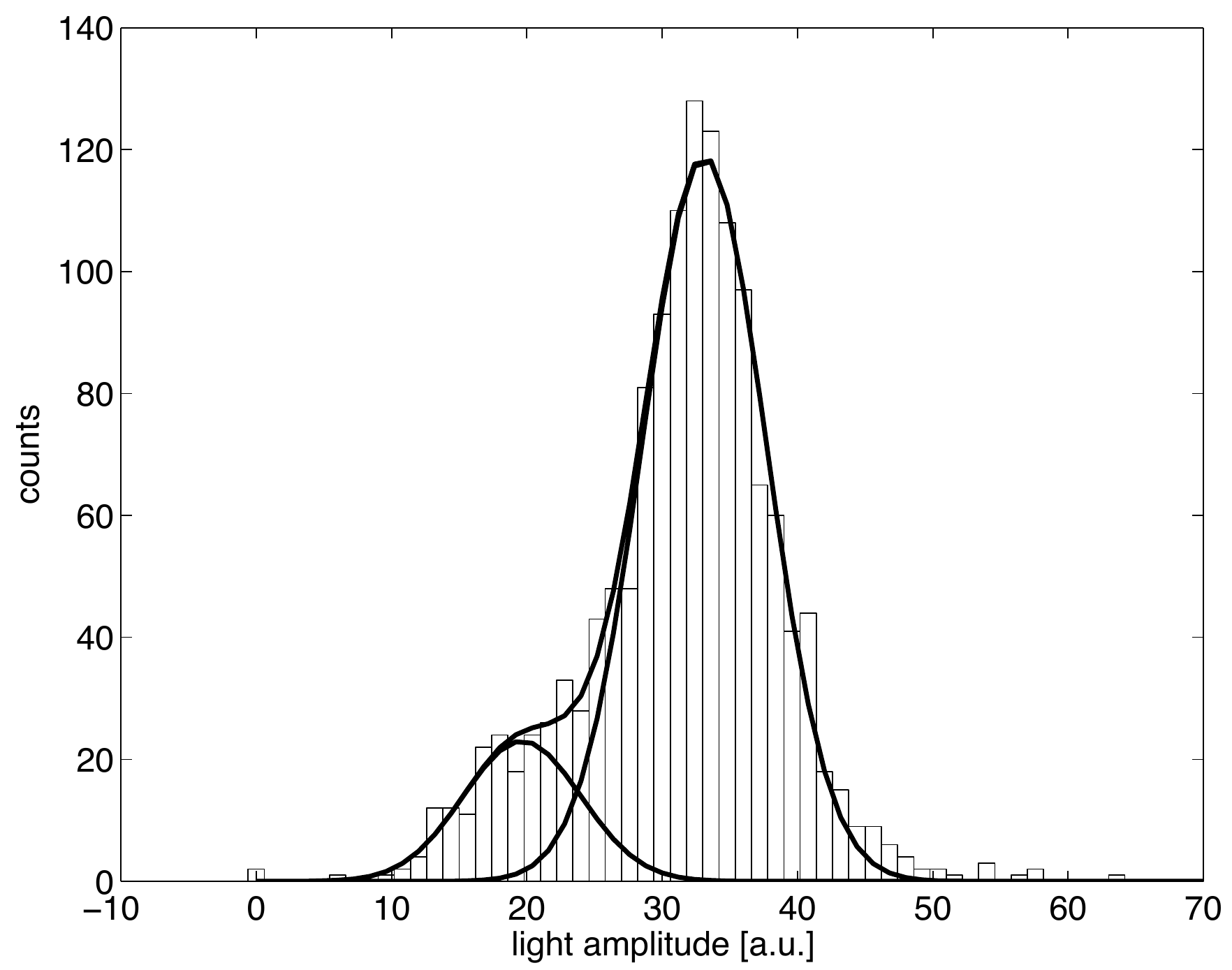} 
  \caption{Histogram of the light amplitude for a section of the nuclear recoil band
(250-280~keV) where only contributions from O and Ca are expected. A two-Gaussian
fit can be performed, in order to obtain values for the quenching factors of the two nuclei.}
  \label{2gauss}
\end{figure}
\end{center}

The events depicted in the histogram of Fig. \ref{2gauss} reflect those marked in grey in the nuclear recoil band between 250 and 280~keV of Fig. \ref{fig: nsf1stbeamtimescatterplot}. For kinematic reasons only oxygen and calcium recoils contribute. A two-Gaussian fit can be performed to obtain the quenching factors of the two nuclei. The results are summarized in Table \ref{tab_QF2}.

The three and two-Gaussian fits describe the corresponding histograms very well. The errors given in Table \ref{tab_QF1} and Table \ref{tab_QF2} include statistical errors only.

The values for the QFs of oxygen and calcium obtained with this measurement are compatible with the values obtained with other methods (however at higher temperatures) as described in detail in \cite{bavykina}. The preliminary value for the QF of tungsten exhibits a significant error. However, its average value fits well the results obtained at higher temperatures in \cite{bavykina}. The error can be further decreased by applying the TOF method as depicted in Fig. \ref{fig: nsfsetup}, which reduces parasitic background and results in an even clearer separation of the individual peaks \cite{jagemann2006}.

\begin{center}
\begin{table}
 \begin{tabular}{c|clc}
Element & Quenching factor (160-200~keV)     \\
\hline
Oxygen   & 7.81 $\pm$ 0.05  \\
\hline
Calcium  & 15.7 $\pm$ 0.8  \\
\hline
Tungsten & 39 +18 -10   \\
 \end{tabular} \\
 \caption{Values of the quenching factors (160-200~keV) obtained from the fits shown in Fig.~\ref{3gauss}.}
\label{tab_QF1}
\end{table}
\end{center}

\begin{center}
\begin{table}
 \begin{tabular}{c|clc}
Element & Quenching factor  (250-280~keV)   \\
\hline
Oxygen   & 8.00 $\pm$ 0.07   \\
\hline
Calcium  & 13.5 +1.1 -0.9  \\
\hline
Tungsten & -  \\
 \end{tabular} \\
 \caption{Values of the quenching factors (250-280~keV) obtained from the fits shown in Fig.~\ref{2gauss}.}
\label{tab_QF2}
\end{table}
\end{center}

\section{Conclusions}\label{conclusions}

A unique low-temperature neutron scattering facility has been set up at the  Maier-Leibnitz-Accelerator Laboratory in Garching. The aim is to perform Quenching Factor measurements at the operating temperature ($\sim$10 mK) of $\mathrm{CaWO_4}$-based CRESST detectors. During the present commissioning activities, which comprised the setup of a $^3$He/$^4$He dilution refrigerator, as well as the development of a dedicated low-temperature detector for high count rates, first results have been obtained using a continuous monoenergetic neutron beam of 11~MeV. In comparison to common neutron sources the higher energy of the neutron beam makes the investigation of tungsten recoils easier.

The results of these preliminary measurements fit well into data published so far on $\mathrm{CaWO_4}$ quenching factors at higher temperatures. 
Once the setup will have been equipped with new readout electronics and additional neutron detectors, this facility will be fully operational. With a view to future multi-target dark matter experiments like EURECA, this neutron characterization facility will be a powerful tool and will allow investigations also on further target materials such as  $\mathrm{ZnWO_4}$, $\mathrm{PbWO_4}$ and others.

\section{Acknowledgements}\label{acknowledgements}

This article is dedicated to the memory of our friend and colleague Wolfgang Westphal, who has so significantly contributed to the success of all our CRESST and EURECA related activities. \\

We would like to thank the whole team of operators and technicians at the Maier-Leibnitz-Accelerator Laboratory in Garching for their valuable support and constant help.\
This work has been supported by funds of the DFG (SFB 375, Transregio 27: ÓNeutrinos and BeyondÓ), the Munich Cluster of Excellence (ÓOrigin and Structure of 
the UniverseÓ), the EU networks for Cryogenic Detectors (ERB-FMRXCT980167) 
and for Applied Cryogenic Detectors (HPRN-CT2002-00322), and the Maier-Leibnitz- 
Laboratorium (Garching).

\end{document}